# A New Chaos-Based Cryptosystem for Secure Transmitted Images

Abir AWAD

**Abstract—**This paper presents a novel and robust chaos-based cryptosystem for secure transmitted images and four other versions. In the proposed block encryption/decryption algorithm, a 2D chaotic map is used to shuffle the image pixel positions. Then, substitution (confusion) and permutation (diffusion) operations on every block, with multiple rounds, are combined using two perturbed chaotic PWLCM maps. The perturbing orbit technique improves the statistical properties of encrypted images. The obtained error propagation in various standard cipher block modes demonstrates that the proposed cryptosystem is suitable to transmit cipher data over a corrupted digital channel. Finally, to quantify the security level of the proposed cryptosystem, many tests are performed and experimental results show that the suggested cryptosystem has a high security level.

## 1 Introduction

SECURE transmission of confidential digital messages has become a common interest in both research and applications. Traditional symmetric ciphers such as Advanced Encryption Standard (AES) are designed with good confusion and diffusion properties. These two properties can also be found in chaotic systems which have desirable properties of pseudo-randomness, ergodicity, high sensitivity to initial conditions and parameters. Chaotic maps have demonstrated great potential for information security, especially image encryption, while the standard encryption methods as the AES algorithm seem not to be suitable to cipher such type of data. The author in [1] for example, shows that the images encrypted by the AES algorithm are still intelligible and then proposes a modified AES to solve this problem. Since the 1990s, a large amount of work using digital chaotic techniques to construct cryptosystems has been studied and has attracted more and more attention in the last years [2], [3], [4], [5], [6], [7], [8]. Researchers are especially interested in enhancing the chaotic generators and the diffusion stage of the cryptosystems. In order to be used in every application, chaotic sequences must seem absolutely random and have good cryptographic properties. Many studies on chaotic maps are drawn [9], [10] and [11]. The researchers in [7], [12] proved that logistic map, that was widely used in the encryption domain, is not enough random and uniform. Then, they propose to use other chaotic maps like Piece Wise Linear Chaotic Map (PWLCM). In [13] and [14], we studied and improved some existing chaotic techniques like PWLCM. The improved chaotic maps generate chaotic signals with desired statistical properties and in compliance with NIST statistical tests. Indeed, to obtain better dynamical statistical properties and to avoid the dynamical degradation caused by the digital chaotic system working in a finite state, a perturbation technique is used. In this paper, we propose a new encryption algorithm controlled by this perturbed PWLCM and then prove its contribution.

Chaotic output signals, which present random statistical properties, are used for both confusion and diffusion operations in a cryptosystem. Confusion obscures the relationship between the plaintext and the ciphertext and diffusion dissipates the redundancy in the plain text by spreading it out over the cipher text.

It is well known that images are different from texts in many aspects, such as high redundancy and correlation. The main obstacle in designing effective image encryption algorithms is the difficulty of shuffling and diffusing such image data by traditional cryptographic means [15], [16], [17], [18]. In most of the natural images, the value of any given pixel can be reasonably predicted from the values of its neighbors.

For image encryption, two-dimensional (2D) chaotic maps are naturally employed as the image can be considered as a 2D array of pixels.

Many researchers have proposed schemes with combinational permutation techniques [19], [20] that divide the image into blocks then shuffle their positions before passing them to the bit manipulation stage. In fact, bit level permutations are particularly difficult for processors. Many researchers tend to avoid them in the design of cryptographic algorithms or use very simple permutations [21], [22]. But recently, a number of candidate instructions have been proposed to efficiently compute arbitrary bit permutations [7], [23], [24] and [25]. In this paper, we propose a new approach for image encryption using a combination of different permutation techniques: Pixel and bit permutations. This combination is interesting because it reduces the correlation between

Abir Awad was with the Operational Cryptology and Virology Laboratory (C+V)^O, ESIEA-OUEST. Drs Calmette et Guérin street, 53003 Laval Cedex, France. And currently, she is with LIFO, University of Orléans, Léonard de Vinci street, 45000 Orléans, France (phone: + 33-2-38417288; e-mail: abir.awad@gmail.com).

the adjacent pixels and enhances the statistical properties of the encrypted images.

Moreover, cryptographic modes for block ciphers have received much attention lately, partly due to the recommendation of the National Institute of Standards and Technology (NIST) [26]. No block cipher is ideally suited for all applications. This comes from differing tolerances of applications to properties of various cryptographic modes. As we search to meet the requirements of secure image transfer, we examine the problem of error propagation in various cipher block modes. And we prove that our algorithm, unlike existing algorithms such as the Socek, Yang, Lian and Wong methods [7], [27], [28], [29], is suitable for the transmission of encrypted images over an insecure channel.

The paper is organized as follows: Section 2 briefly introduces the original schemes proposed by Socek [7], Yang [27], Lian [28] and Wong [29]. Section 3 describes the proposed algorithm; Section 4 introduces the perturbed chaotic map used; Section 5 explains the substitution and permutation transformations used in the algorithm; Section 6 shows the decryption process. Section 7 presents the cipher block modes used for the encryption and compares the theoretical propagation error induced by each mode. The simulation results and security analysis are given in section 8. And finally, we summarize our conclusions in section 9.

## 2 Overview of Three Existing Algorithms

### 2.1 Enhanced 1-D Chaotic Key Based Algorithm for Image Encryption

The ECKBA [7] encryption algorithm (Fig.1) transforms an image I using a substitution (S) and a bit permutation (P) network controlled by a PWLCM chaotic map. The algorithm performs r rounds of the SP-network on each pixel. The next iteration i+1 of the chaotic map is perturbed using the previous cipher block $c_i$.

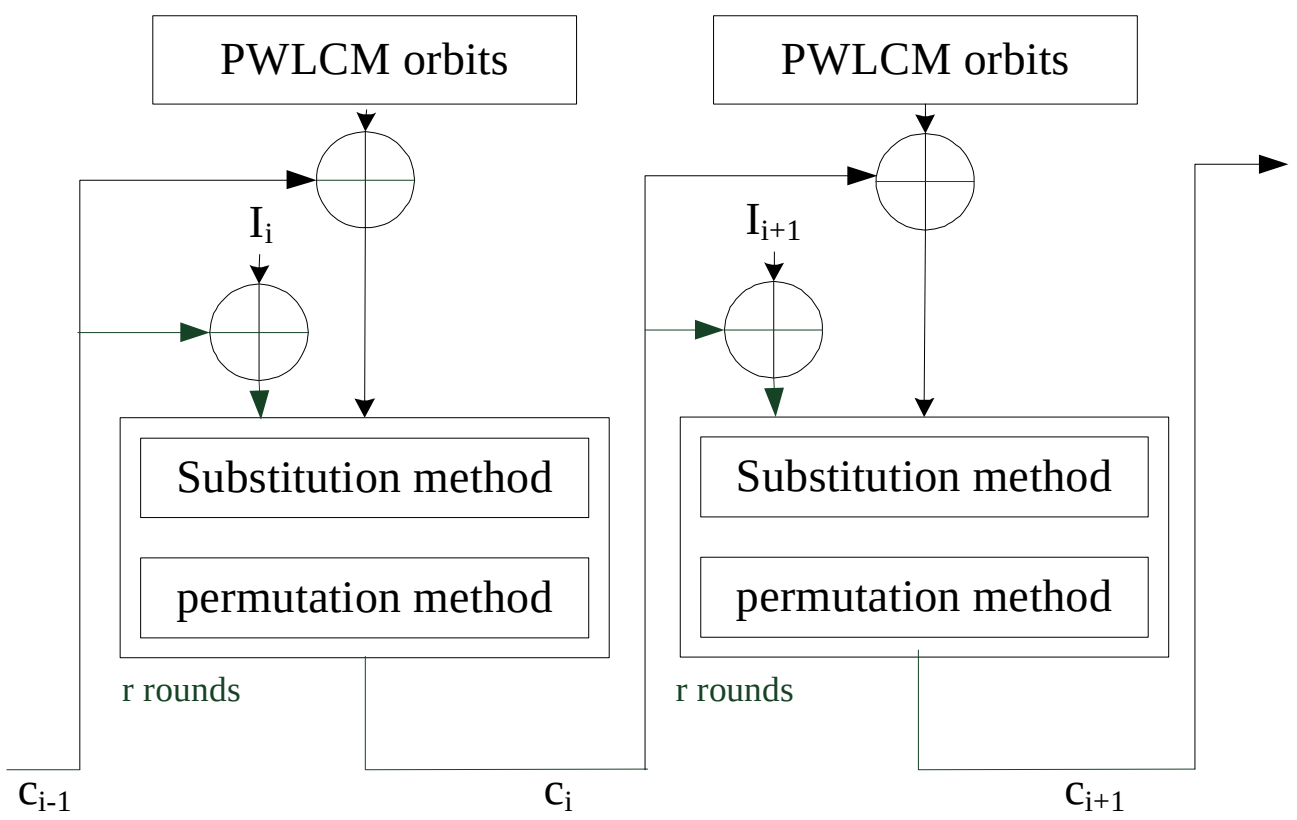

Fig. 1. ECKBA encryption algorithm

In addition to this, the algorithm implements a cipher-block chaining (CBC) encryption mode. Each block of plaintext $I_{i+1}$ is XORed with the previous ciphertext block $c_i$ before being encrypted. The permutation and the substitution methods are explained in section 5.

### 2.2 Yang encryption algorithm

In his paper [27], Yang proposed a novel method for encryption based on iterating map with output-feedback. The output feedback, instead of simply mixing the chaotic signal of the proposed chaotic cryptosystem with the cipher-text, is relating to previous cipher-text that is obtained through the plaintext.

He used a simple one-dimensional logistic map defined as follows:

$$x(n) = b\,x(n-1)(1-x(n-1)) \quad (1)$$

where the positive control parameter $b$ and the chaotic value x(i), i=1, 2,…, n, are real values and respectively belong to the intervals (3.58; 4) and (0; 1).

Provided that the current chaotic state is x(n) and its binary representation is

$$x(n) = 0.x_1(n)x_2(n)...x_i(n)...x_N(n) \qquad x_i(n) \in \{0,1\} \quad i = 1,2,...,N \quad (2)$$

By defining three variables whose binary representation is $x_h^{'} = x_1(n)...x_{15}(n)$, $x_p^{'} = x_{16}(n)...x_{30}(n)$, $x_s^{'} = x_{31}(n)...x_{45}(n)$ respectively. We can define the following equations:

$$f(x(n)) = x_h^{'} \oplus x_p^{'} \oplus x_s^{'} \quad (3)$$

$$g(x(n)) = f(x(1)) \oplus ... \oplus f(x(n)) \quad (4)$$

Also, a value called $Outfd$ is given. Its value is the bitwise exclusive-OR (XOR) operation of previous eight cipher-texts.

$$Outfd = c_{i-8} \oplus c_{i-7} \oplus ... \oplus c_{i-1} \quad (5)$$

The cipher-text is obtained by the bitwise exclusive-OR (XOR) operation with sequences I′(n) and Aj. I′(n) is the permuted value of the plaintext block with a cyclic shift operation [30]. The value of the variables Aj is obtained by the method of Fig. 2.

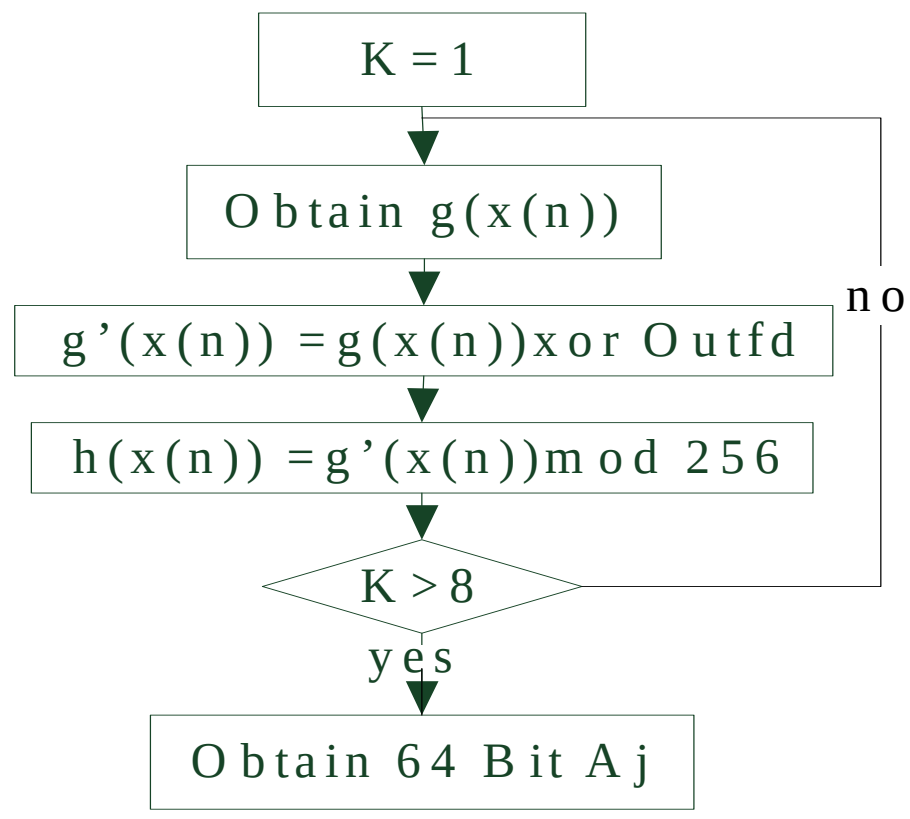

Fig. 2. Yang encryption algorithm

### 2.3 Lian and Wong encryption algorithms

In [28], Lian suggested using a standard map for diffusion while keeping the logistic map for pixel value confusion. The diffusion stage permutes the pixels in the image, without changing its value. In the confusion stage, the pixel values are modified (Fig.3).

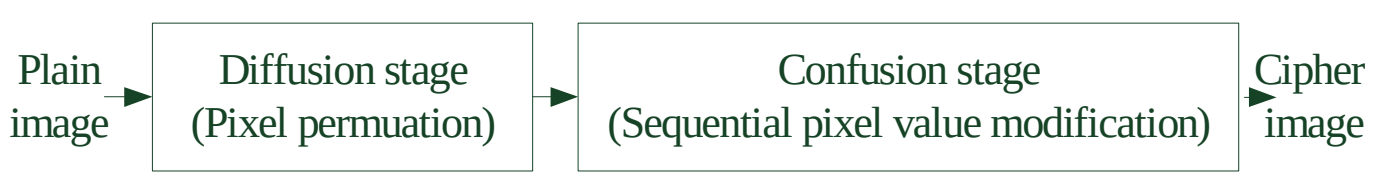

Fig. 3. Combinational encryption algorithm

In the Lian method, the diffusion process is achieved by using a 2D standard map and the confusion stage is achieved by eq. (6).

$$c_i = p_i \oplus f(c_{i-1}) \quad (6)$$

Where $p_i$ is the value of the i^th block of the permuted image, $c_{i-1}$ and $c_i$ are respectively the values of the (i-1)^th and the i^th pixels of the confused image. *f(.)* is a non linear function.

As the new pixel value is obtained by an exclusive OR operation, the inverse operation can be made as follows (eq. (7)):

$$p_i = c_i \oplus f(c_{i-1}) \quad (7)$$

Where $p_i$ is the value of the i^th block of the plain image, $c_{i-1}$ and $c_i$ are respectively the values of the (i-1)^th and the i^th pixels of the ciphered image.

The Wong scheme [29] is a modification of the one suggested by Lian. The pixel value mixing effect of the whole cryptosystem is contributed by two operations: a modified diffusion process and the original confusion function. In the modified diffusion stage, the new position of a pixel is calculated using the standard map. Before performing the pixel relocation, confusion effect is injected by adding the current pixel value of the plain image to the previous permuted pixel and then performs a cyclic shift.

## 3 ENCRYPTION ALGORITHM

In this section, we present the developed block encryption algorithm of images called CBCSTI that we implemented with Matlab.

Let *I* be an *MxN* image with b-byte pixel values, where a pixel value is denoted by *I(i)*, $0 \le i < MxNx3$. A block cipher is an encryption scheme which breaks up the plaintext messages into blocks of a fixed length and encrypts one block at a time.

The algorithm characteristics and steps are (see Fig. 4):

1. The key size is 128-bits.

2. The chaotic maps used are a 2 D chaotic map and two perturbed PWLCM.
   Firstly, the positions of the pixels of the original image are shuffled by 2D chaotic map.

   (xn,yn)=2D chaotic map(x,y)

   (x,y) is the initial pixel position and (xn,yn) presents the novel pixel position.

   Then, the perturbed chaotic values are generated each r iterations for r/4 times to control the substitution method and r/4 or r/2 times to control the bit permutation technique.

3. A substitution box (S-box) is applied.

   O(i)=substitution(O(i), c(mod((i+j),r))

   O(i) can be the plaintext block or the output key that we encrypted. It depends on the adopted operation mode. c(mod((i+j),r)) is the chaotic value that we used to control the substitution technique. We transform the perturbed chaotic value into a binary value to control the substitution technique. These bits are decomposed on four bytes and then transformed on four integers c(1), c(2), c(3) and c(4). These values are used for the r rounds of the substitution.
   This operation is repeated for the r blocks O(i)…O(i+r-1).
   The chaotic values are generated r/4 times each r iterations to insure the required chaotic blocks. And, these chaotic blocks c(k) for k=1…4 are not applied in the same order on each of the corresponding plaintext or output key blocks.

4. A bit permutation box (P-box) adding diffusion to the system is performed.
   The bits of each pixel value are shuffled using a bit permutation method: Socek or Cross method. The used bit permutation method is controlled by the second perturbed PWLCM map.

   O(i)=permutation(O(i), d(mod((i+j),r))

   O(i) is the substituted block that we permute and d(mod((i+j),r) is the controlling chaotic block. Similar to the substitution technique, the chaotic binary value is decomposed on two parts of 16 bits if we choose the Socek method and 8 bits if the Cross method is used. These binary blocks are then transformed onto decimal values. Then, the second Perturbed PWLCM is performed r/2 or r/4 times each r iterations depending on the chosen bit permutation method.

5. Multiple rounds r for encryption and decryption processes are used.

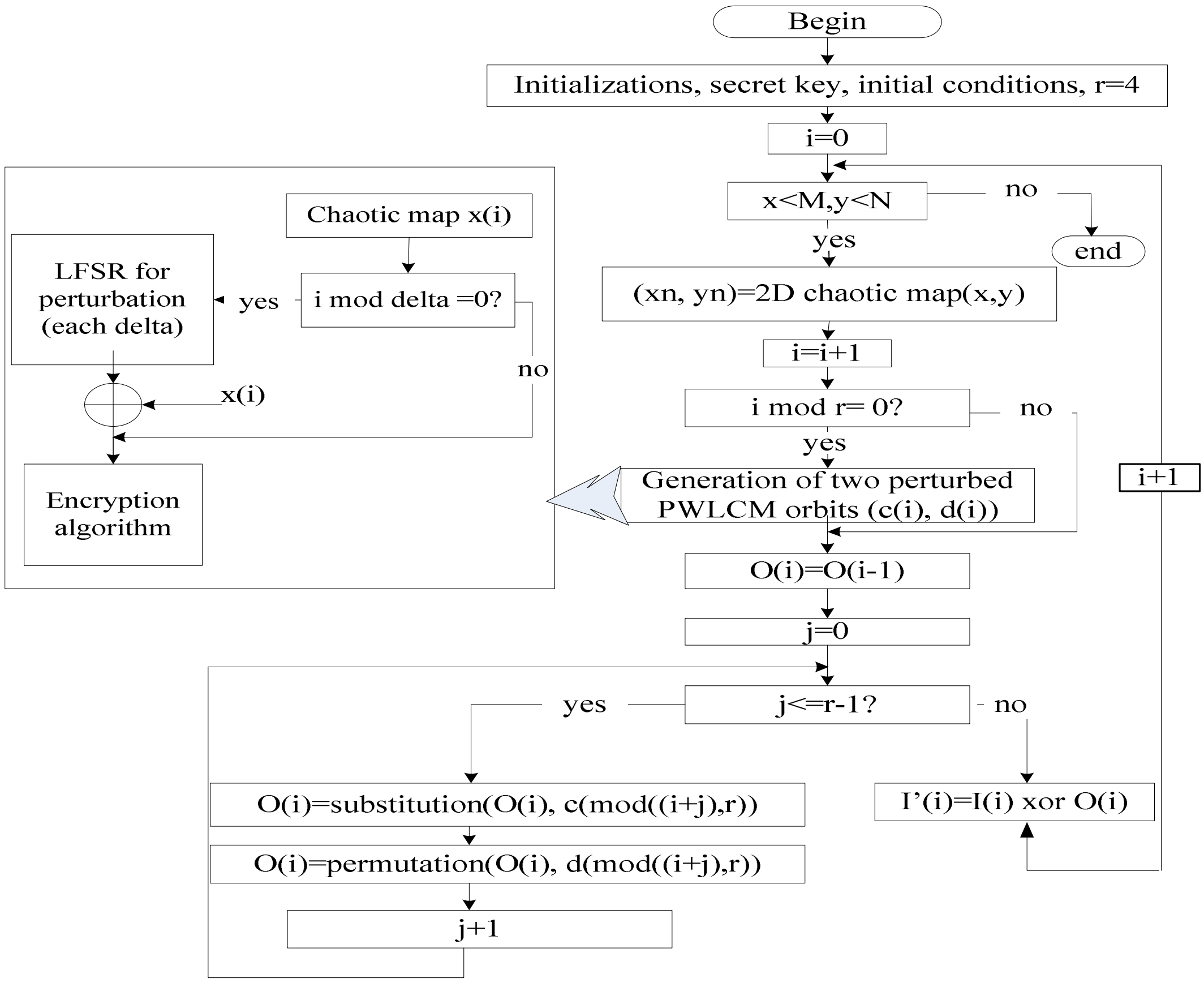

Fig. 4. The proposed algorithm in the OFB operation mode

The encryption algorithm transforms an image I using a 2D chaotic map and an S-P network generated by a one dimensional chaotic map and a 128-bit secret key. The algorithm performs r rounds of an SP-network on each pixel. Fig. 4 illustrates the flow chart of the algorithm with Output Feedback operation mode (OFB).

Many variants of the algorithm are drawn. They differ depending on the choice of pixel and bit permutation methods:

CBCSTI-A: uses the standard map to do the pixel position permutation and the Socek method as a bit permutation method.

CBCSTI-B: uses the Standard and CROSS methods.

CBCSTI-C: uses the Arnold and Socek methods.

CBCSTI-D: uses the Arnold and CROSS methods.

CBCSTI-E: uses the Socek method without a 2D chaotic map.

In the next section, we explain the perturbed PWLCM map used in the algorithm. Then, we discuss the SP box adopted in the proposed algorithm.

# 4 PERTURBED PWLCM MAP

## 4.1 PWLCM map

Due to the poor balance property of the Logistic map, some implementations use the following PWLCM map with better balance property [7], [12].

A piecewise linear chaotic map (PWLCM) is a map composed of multiple linear segments.

$$x(n) = F[x(n-1)] = \begin{cases} x(n-1)\times\dfrac{1}{p} & \text{if } 0 \le x(n-1) < p \\ [x(n-1)-p]\times\dfrac{1}{0.5-p} & \text{if } p \le x(n-1) < 0.5 \\ F[1-x(n-1)] & \text{if } 0.5 \le x(n-1) < 1 \end{cases} \qquad (8)$$

where the positive control parameter $p$ and the chaotic value $x$(i), i=1, 2,…, n, are real values and respectively belong to the intervals (0; 0.5) and (0; 1).

## 4.2 The Periodicity of chaotic orbit

Since digital chaotic iterations are constrained in a discrete space with $2^N$ elements, it is obvious that every chaotic orbit will eventually be periodic and will finally go to a cycle with a limited length not greater than $2^N$ [31], [32]. Generally, each digital chaotic orbit includes two connected parts: $x(1), x(2),..., x(l)$ and $x(l), x(l+1),..., x(n)$, which are respectively called "transient branch" and "cycle". Accordingly, $l$ and $n+1$ are respectively called "transient length" and "cycle period", and $l+n$ is called "orbit length". Then, some questions might arise: how to estimate the transient lengths and cycle periods? Are the lengths large enough to ensure the dynamical properties of continuous chaotic systems? Unfortunately, as B.V.

Chirkikov and F. Vivaldi demonstrated in [33], rigorous studies of such estimations (especially the average lengths) are notoriously difficult and the difficulties are actually from the lack of an ergodic theory for discrete chaotic systems. Since the theoretical analysis is not possible, statistical experiments are widely used to explore this issue.

### 4.3 Perturbed chaotic orbit

To improve the dynamical statistical properties of generated chaotic sequences, a perturbation-based algorithm is used. The cycle length is expanded and consequently good statistical properties are reached. Many perturbation techniques are proposed. For example, Socek [7] and Yang [27] use a perturbation-based algorithm. The orbits are perturbed by the encrypted blocks. The Socek algorithm is very secure but a bit error transmission causes a random number of erroneous bits in the decrypted image. In this paper, we use another perturbation technique using maximal length LFSR, which is a suitable candidate for perturbing the signal generator [31].

The fundamental basis of the perturbing method consists in the fact that the chaotic system runs away from the cycle, i.e. the chaotic system having entered a cycle can be driven to leave the cycle immediately by a perturbance. We choose to perturb the chaotic orbit by the maximal length LFSR because its generated sequences have the following advantages: controllable long cycle length *($2^k$- 1)* (*k* is the polynomial degree); uniform distribution and delta like autocorrelation function. These characteristics ensure that the good statistical properties of chaos dynamics will not be degraded.

The perturbing bit sequence can be generated every *n* clock as follows:

$$Q_{k-1}^{+}(n) = Q_k(n) = g_0 Q_0(n) \oplus g_1 Q_1(n) \oplus \ldots \oplus g_{k-1} Q_{k-1}(n) \quad with\ n = 0,1,2,\ldots \tag{9}$$

Where $\oplus$ represents 'exclusive or', $g = [g_0\ g_1 \ldots g_{k-1}]$ is the tap sequence of the primitive polynomial generator, and $Q_0\ Q_1 \ldots Q_{k-1}$ are the initial register values of which at least one is non zero.

The perturbation begins at *n= 0*, and the next ones occur periodically every $\Delta$iterations ($\Delta$ is a positive integer), with *n=* $l \times \Delta$, *l=1,2,…,* The perturbed sequence is given by the equation (10):

$$x_i(n) = \begin{cases} F[x_i(n-1)] & 1 \le i \le N-k \\ F[x_i(n-1)] \oplus Q_{N-i}(n) & N-k+1 \le i \le N \end{cases} \tag{10}$$

Where $x_i(n)$ and $F[x_i(n)]$ represent respectively the *i*th bits of $x(n)$ and $F[x(n)]$.

The perturbation is applied on the last *k* bits of $F[x(n)]$.

When $n \ne l \times \Delta$, no perturbation occurs, so $x(n) = F[x(n-1)]$.

The lower boundary of the system cycle length is

$$T_{\min} = \Delta \times \left(2^k - 1\right) \tag{11}$$

## 5. CHAOTIC PERMUTATION AND SUBSTUTION METHODS

In this section, we present the different chaotic substitution and permutation methods used in the proposed algorithm. This paper compares the contribution of these methods in the whole algorithm.

In the proposed algorithm, we used two perturbed PWLCM chaotic maps. The chaotic value *x(i)* is in the interval (0; 1) (see eq. (8)). Then, a discretization method is applied to transform it to a fixed 32 bit float using the following equation (see eq. (12)):

$$y(i) = round(x(i).2^{32}) / 2^{32} \tag{12}$$

where *x(i)* is a chaotic real value and *y(i)* is the discretized one. The *round* function (instead of *floor* and *ceil*) insures the minimal degradation of the chaotic map. The advantage of this function is discussed in [34].

The proposed perturbation technique is then applied on *y(i)* that is subsequently used to control the S-P box (the substitution and bit permutation operations).

### 5.1 Substitution box

A complex substitution box is applied (eq. 13).

$$\mathrm{Sigma}_r(u,v) = \begin{cases} u \oplus v & if\ r\ is\ even \\ (u+v) \bmod 256 & if\ r\ is\ odd \end{cases} \tag{13}$$

where *u* and *v* are two bytes and *r* is the round value.

### 5.2 Permutation box

In order to disturb the high correlation among adjacent pixels, we propose a scheme that includes two phases: firstly, the pixel positions are permuted by 2D chaotic map. Then, the gray values of the permuted image are encrypted by a bit permutation method.

#### 5.2.1 Permutation of pixel position

The 2D chaotic map shuffles the pixel positions of the plain image and then disturbs the high correlation among pixels. Without loss of generality, the size of the original grayscale image *I* is assumed NxN.

**Permutation by the Arnold cat map**

The coordinates of the pixel positions are $S = \{(x,y) | x,y = 0,1,2,\ldots,N-1\}$. Arnold cat map [20] is described as follows (eq. 14).

$$\begin{bmatrix} x' \\ y' \end{bmatrix} = A \begin{bmatrix} x \\ y \end{bmatrix} (\bmod N) = \begin{bmatrix} 1 & t \\ q & tq+1 \end{bmatrix} \begin{bmatrix} x \\ y \end{bmatrix} (\bmod N) \tag{14}$$

Where $p$ and $q$ are positive integers, $det(A)=1$. The $(x, y)$ and $(x', y')$ are the original and the new positions, respectively. After several iterations, the original image can be permuted completely. The parameters $t$, $q$ and the iteration number $M$ can be chosen as the secret keys.

**Permutation by Standard map**

The Standard map is described with the following formula (eq. 15):

$$\begin{cases} x' = (x+y) \bmod N \\ y' = (y + k\sin(\frac{xN}{2\pi})) \bmod N \end{cases} \quad (15)$$

Where $k$ is a positive constant, $(x, y)$ is the original pixel position and $(x', y')$ is the new one.

### 5.2.2 Bit permutation method

The permutation is made on the bits of each block formed by a byte. In other words, we use a permutation of degree 8 to add diffusion to the system. Actually, the fastest way to achieve this is by using a look up table approach. This approach is fast but the memory requirements are considerably high. A number of permutation methods have been proposed [7], [23], [24] and [25]. Among these, the Cross and Socek methods are the most attractive. They have good cryptographic properties [35].

**Cross permutation**

The Cross method is based on the Benes network, which is formed by connecting two butterfly networks of the same size back-to-back [25]. Cross instruction is defined as follows:

*R3=Cross( m1, m2, R1, R2)*

*R1* is the source register which contains the bits to be permuted, *R2* is the configuration register and *R3* is the destination register for the permuted bits.

As we said, we propose to control this method using chaotic values. Then, in each iteration, the control array *R2* is filled by a chaotic binary suite (8 bits). One Cross instruction performs two basic operations on the source according to the contents of the configuration register and the values of *m1* and *m2*. Fig. 5 shows how the Cross instruction works on 8-bit systems.

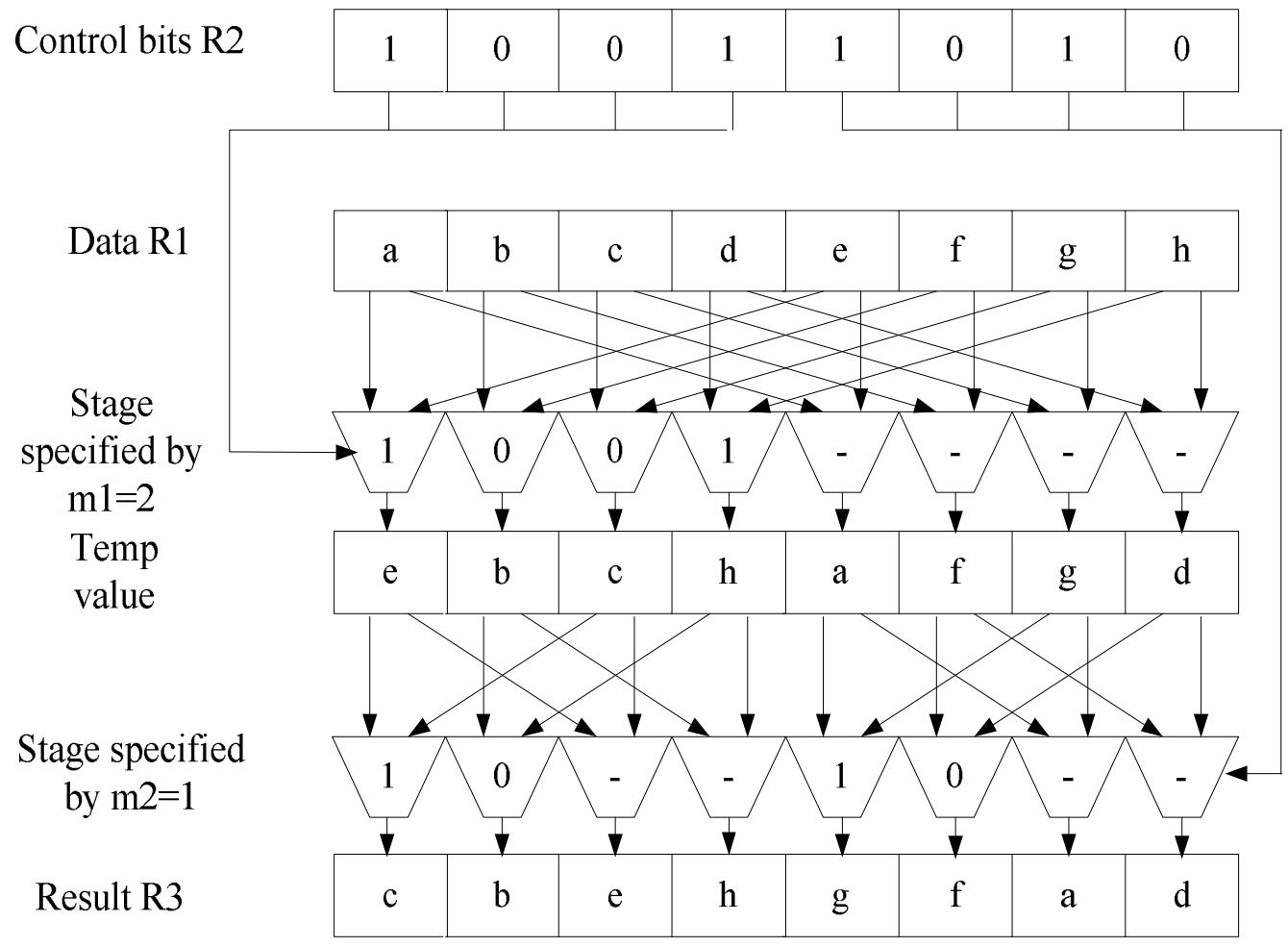

Fig.5. The Cross instruction on 8-bit systems

**Socek permutation**

The permutation method proposed by Socek [7] is a computational approach of degree 8. It permutes the indices of bits of each pixel using the chaotic value. These indices are placed in an array *pj= [1, 2, 3, 4, 5, 6, 7, 8]* and we have then to permute the elements of this array using the chaotic value. Then, the bits are rearranged according to the permuted indices of the new array *P'j*. Fig. 6 presents an example of the Socek method applied on 8 bits. In this case, the new array of indices obtained is *P'j=[4, 6, 7, 1, 3, 8, 2, 5]*.

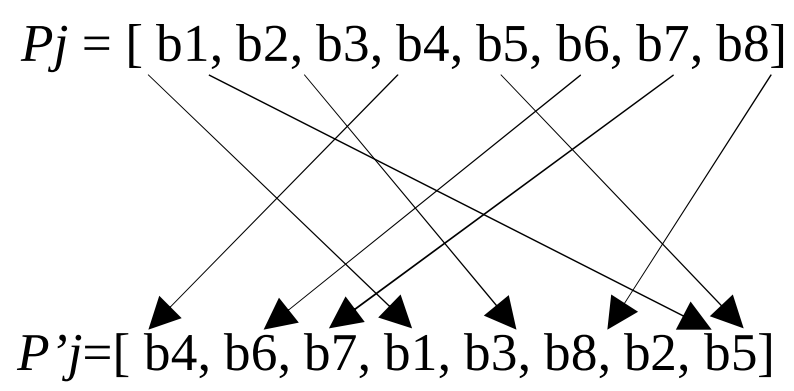

Fig. 6. Socek method on 8 bits

## 6 DECRYPTION PROCESS

The decryption algorithm depends on the cipher mode used. For the modes OFB, CFB and CTR, the decryption algorithm is the same as that of the encryption. But for the CBC mode, it differs slightly from the encryption algorithm. To decrypt an encrypted image, we need to perform the inverse transformations. The inverse substitution will be as follows (eq. 16):

$$\mathrm{Sigma}_r^{-1}(u,v) = \begin{cases} u \oplus v & \text{if } r \text{ is even} \\ (u-v) \bmod 256 & \text{if } r \text{ is odd} \end{cases} \quad (16)$$

The inverse CROSS instruction is the same as that used for the encryption process but the contents of the configuration register *m1* and *m2* are exchanged. And, in the inverse Socek method, the bits are rearranged according to the array indices *(8-p(i))* instead of *p(i)* used in the encryption process. Therefore, we need to reverse the order of the substitution and bit permutation methods. Then, we use the inverse methods to decrypt the image.

## 7 CRYPTOGRAPHY MODES AND ERROR PROPAGATION

A cryptographic mode usually combines the basic cipher, some sort of feedback, and some simple operations. Some applications need to parallelize encryption or decryption, while others need to be able to preprocess as much as possible.

A bit error is the substitution of a '0' bit for a '1' bit, or vice versa. These errors are generated by the transmission channel as a consequence of interference and noise.

The error propagation phenomenon implies that errors in the encrypted text produce errors in the decrypted plaintext. So, it is important that the decrypting process be able to recover from bit errors in the ciphertext.

In this section, we examine the problem of error propagation in various cipher block modes of operation, such as: Cipher Block Chaining (CBC), Cipher Feedback (CFB), Output Feedback (OFB), and Counter mode (CTR) [26]. The dependence between input and output error probability of the modes is presented. The results obtained can be used to choose the block cipher and its mode to generate a suitable cryptogram for transmission over a noisy channel.

The CFB is a special mode with which segments are operated. The segment is an *s* bit block, where $1 \le s \le b$. The j-th plaintext and encrypted segment are denoted by $I_j^{\#}$ and $I_j^{'\#}$ respectively.

The effect of the error bit $I_{i,j}^{'}$ or $I_{i,j}^{'\#}$ in the block $I_j^{'} = \left(I_{1,j}^{'}, I_{2,j}^{'}, ..., I_{b,j}^{'}\right)$ or $I_j^{'\#} = \left(I_{1,j}^{'\#}, I_{2,j}^{'\#}, ..., I_{b,j}^{'\#}\right)$ on the appearance of errors in the plaintext for individual modes is summarized in the following table (table I).

TABLE 1
THE EFFECT OF BIT ERRORS FOR CIPHER BLOCK MODES

| MODE | Effect of bit errors in $I_j^{'}$ |
|---|---|
| ECB | RBE in $I_j$ |
| CBC | RBE in $I_j$<br>SBE in $I_{j+1}$ |
| CFB | SBE in $I_j^{\#}$<br>RBE in $I_{j+1}^{\#}$, $I_{j+2}^{\#}$,..., $I_{j+b/s}^{\#}$ |
| OFB | SBE in $I_j$ |
| CTR | SBE in $I_j$ |

In the table, SBE (Specific Bit Errors) means that an individual error bit $I_{i,j}^{'}$ or $I_{i,j}^{'\#}$ produces in the appropriate decrypted block an individual error bit $I_{i,j}$ or $I_{i,j}^{\#}$. It occurs in the same bit positions as the bit errors in the encrypted image. RBE (random bit errors) means that an individual error bit $I_{i,j}$ or $I_{i,j}^{\#}$ randomly affects all bits in the decrypted block $I_{i,j}$ or in the segments $I_{j+1}^{\#}$, $I_{j+2}^{\#}$,..., $I_{j+b/s}^{\#}$.

In the CBC mode for example, all bit positions that contain bit errors in a cipher text block will produce an RBE in the same decrypted block and an SBE in the second one; the other bit positions are not affected. For the OFB and CTR modes, bit errors within a ciphertext block do not affect the decryption of any other block.

If $P_e$ denotes the bit error probability in the decrypted image and $p_e$ is the bit error probability in the encrypted image. As we said, in the case of OFB and CTR modes, only the SBE type of error propagation can occur. For these modes, each error bit $I'_{i,j}$ of the encrypted image causes only one incorrect bit $I_{i,j}$ of the original image and thus the output error probability is equal to the input one:

$$P_e = p_e \tag{17}$$

In the case of other modes (ECB, CBC, CFB), the RBE type of error appears. To present the dependence between the two probabilities $p_e$ and $P_e$, we will give some definitions.

The probability *P(x)* that there are *x* error bits out of *b* received bits, is given by the formula (eq. 18):

$$P(x) = \begin{bmatrix} b \\ x \end{bmatrix} . p_e^x . (1 - p_e)^{b-x}, x = 01, 2, ...b \tag{18}$$

Then, it holds for the probability $P_0$, that *b* bits are correct:

$$P_0 = (1 - p_e)^b \tag{19}$$

and for the probability $Q_0$, that at least one bit is incorrect:

$$Q_0 = 1 - P_0 = 1 - (1 - p_e)^b \tag{20}$$

We call $P_0$ the correct block probability and $Q_0$ the incorrect block probability.

The probability $P_h$ that the output bit changes its value as a consequence of modifying the input block is called the bit inversion probability. We assume that $P_h = 1/2$.

In the ECB mode, the output error probability *Pe* is equal to:

$$P_e = P_h Q_0 = \frac{1}{2}.[1 - (1 - p_e)^b] \tag{21}$$

The resulting output error probability $P_e$ for the CBC mode is given by the equation:

$$P_e = p_e.(1 - p_e)^b + \frac{1}{2}.[1 - (1 - p_e)^b] \tag{22}$$

In fact, $I_j = U_j \oplus I_{j-1}^{'}$ where $U_j = D_k(I_j^{'})$.
The bit $I_{ij}$ is incorrect in the following cases:
a) the bit $I'_{I,(j-1)}$ is incorrect and the block $I'_j$ is correct,

b) the bit $I'_{I,(j-1)}$ is incorrect, the block $I'_j$ is incorrect and the bit $u_{i,j}$ is not inverted,

c) the bit $I'_{I,(j-1)}$ is correct but the block $I_j$ is incorrect and the bit $u_{i,j}$ is inverted.

The probability of the error bit $I'_{I,(j-1)}$ is equal to $p_e$ and the probability of the correct block $I'_j$ is $P_0$. Thus, the situation occurs with the probability $P$a) = $p_e.P_0$. The probability of the incorrect block $I'_j$ is equal to $Q_0$ and the probability that the bit $u_{i,j}$ is not inverted, amounts to (1-$P_h$). Then, the probability of the situation b) is equal to the quantity $P$b) = $p_e.Q_0.(1-P_h)$. The probability of the correct bit $I'_{i,(j-1)}$ is equal to the value (1-$p_e$), the probability of the incorrect block $I'_j$ is equal to $Q_0$ and the probability that the bit $u_{i,j}$ is inverted, amounts to $P_h$. Then, the probability of the situation c) is equal to the quantity $P$c) = $(1-p_e).P_h.Q_0$. Thus the resultant output error probability $Pe$ for the CBC mode is given by this equation:

a)

$$P_e = P_{a)} + P_{b)} + P_{c)} = P_{a)} = p_e.P_0 + \frac{1}{2}Q_0$$
$$= p_e.(1-p_e)^b + \frac{1}{2}.\left[1-(1-p_e)^b\right] \quad (23)$$

The equations in the case of (CBC) and (CFB) are the same. It follows that the output error probability $Pe$ is the same for both of the modes. Thus, the CBC and CFB modes are equivalent from the viewpoint of error propagation. From equality (CFB), it is also evident that the output error probability of the CFB mode does not depend on the length $s$ of segments. Table II presents the dependence between the two probabilities $p_e$ and $P_e$ for different operation modes [36].

TABLE 2
THE DEPENDENCE BETWEEN THE BIT ERROR PROBABILITIES IN THE ENCRYPTED AND DECRYPTED IMAGES

| Mode | Effect of bit errors in $I'_j$ |
|---|---|
| ECB | $P_e = \frac{1}{2}.\left[1-(1-p_e)^b\right]$ |
| CBC, CFB | $P_e = p_e.(1-p_e)^b + \frac{1}{2}.\left[1-(1-p_e)^b\right]$ |
| OFB,CTR | $P_e = p_e$ |

## 8 SIMULATION RESULTS AND SECURITY ANALYSIS

Some experimental results are given in this section to demonstrate the efficiency of our scheme. We implemented the different algorithms in Matlab. The computer used is a Pentium(R) D CPU 3Ghz, 2.99 Ghz, 2 Go RAM. The plain images used are 'MANDRILL.BMP' (Fig. 7(a)) and 'LENA.BMP' (Fig. 7(b)) with the size 512x512x3. The cipher image of Mandrill with CBCSTI-A algorithm is shown in Fig.8. Similar image is found for Lena image and similar results are obtained for the other versions of the proposed CBCSTI algorithm.

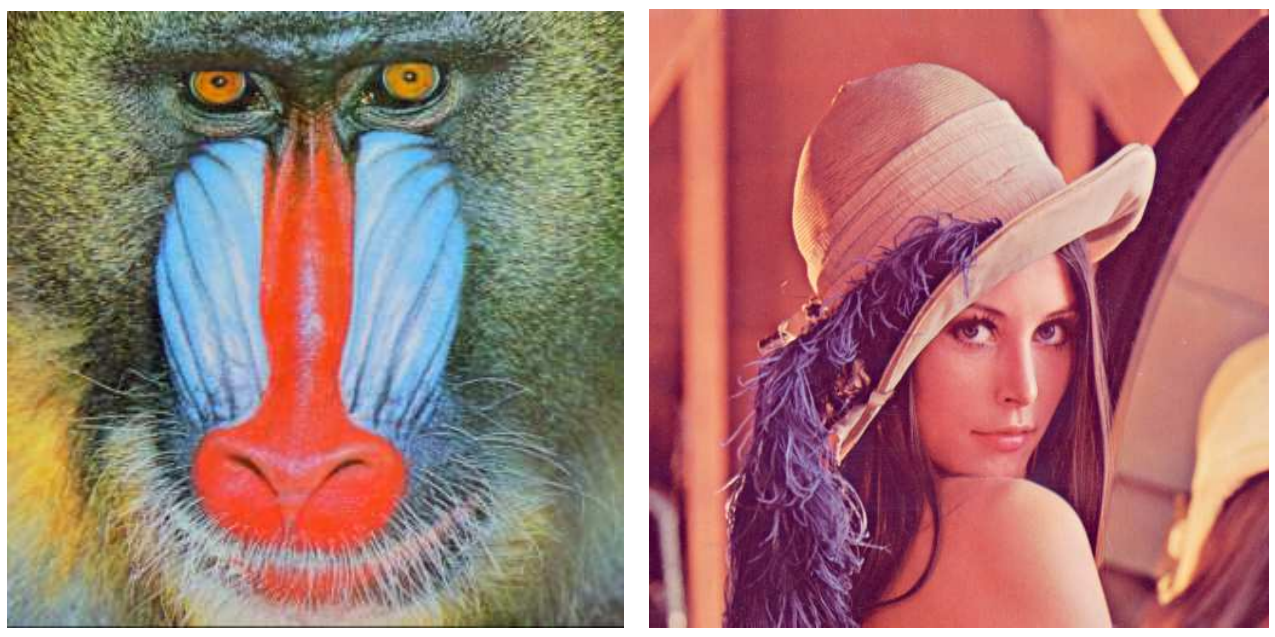

(a) (b)

Fig.7. (a) 'MANDRILL.BMP' image and (b) 'LENA' image

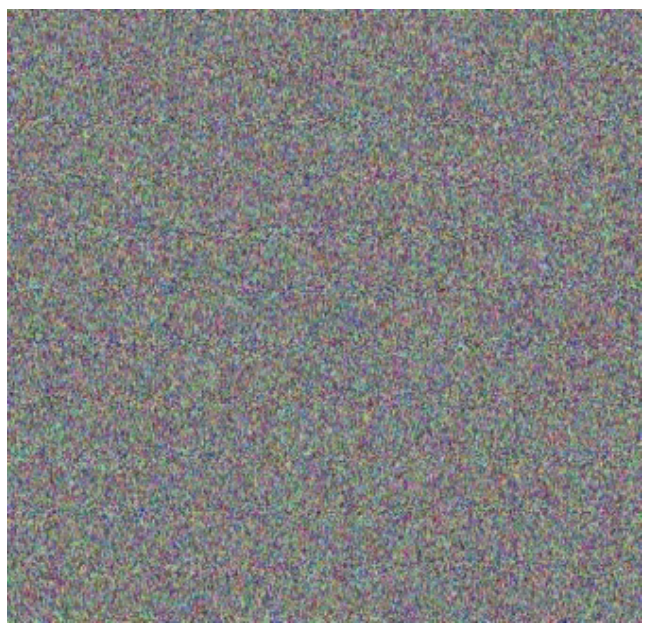

Fig.8. The encrypted image of 'MANDRILL.BMP'

As we can see, the images encrypted by the proposed algorithm are unintelligible. The random quality of the ciphered image is more studied in the following sections.

### 8.1 Comparison between the proposed algorithm and existing chaotic encryption methods

In section 7, we studied, in different operation modes, the error propagation phenomenon in the decrypted plaintext when an error has occurred in the encrypted one.

In first part of this section, we examine the problem of error propagation for the proposed CBCSTI-A algorithm, the AES algorithm and the ECKBA(Enhanced 1-D Chaotic Key Based Algorithm) algorithm [7]. We encrypt the Mandrill and Lena images with the these algorithms and we measure the effect of a random bit error in the encrypted image on the decrypted one. Table 3 presents an example of the performed statistical test in OFB and CBC operation modes. The ECKBA algorithm was proposed in CBC mode. For this reason, we made the comparison between the two algorithms in this mode.

TABLE 3
THE EFFECT OF A TRANSMISSION BIT ERROR ON THE DECRYPTED IMAGES OF MANDRILL AND LENA FOR CBCSTI-A AND ECKBA ALGORITHMS.

| The erroneous blocks in the ciphered image | Number of erroneous blocks in the deciphered image | | | |
|---|---|---|---|---|
| | CBCSTI-A / AES | | ECKBA | |
| | OFB/CTR mode | CBC mode | CBC mode | |
| | Mandrill/Lena | Mandrill/Lena | Mandrill | Lena |
| (120, 17, 1) | 1 | 2 | 105 | 77 |
| (256, 512, 2) | 1 | 2 | 1444 | 499 |
| (100, 10, 3) | 1 | 2 | 5622 | 4625 |

As we can see in table 3, using ECKBA algorithms, a transmission bit error causes a random number of erroneous bits in the decrypted image. But, when we change the same bit in the image encrypted with our algorithm, we always find two erroneous blocks in the decrypted one. For the OFB and CTR modes, bit errors within an encrypted block do not affect the decryption of any other block. Same results are obtained for the standard AES encryption algorithm.

The results obtained for the ECKBA algorithm [7] and the Yang algorithm [27] are not compliant with the recommendations exposed in [26]. In fact, in their algorithms, they use a perturbation technique of the chaotic map using the encrypted data (Fig.9(a)). Then, if a transmission error occurs in the ciphered image, we obtain random errors in the decrypted image.

However, in our algorithm, we perturb the chaotic value with an LFSR (Fig.9(b)). The encrypted blocks are independent. As a result, we manage to avoid the propagation error in the decrypted image.

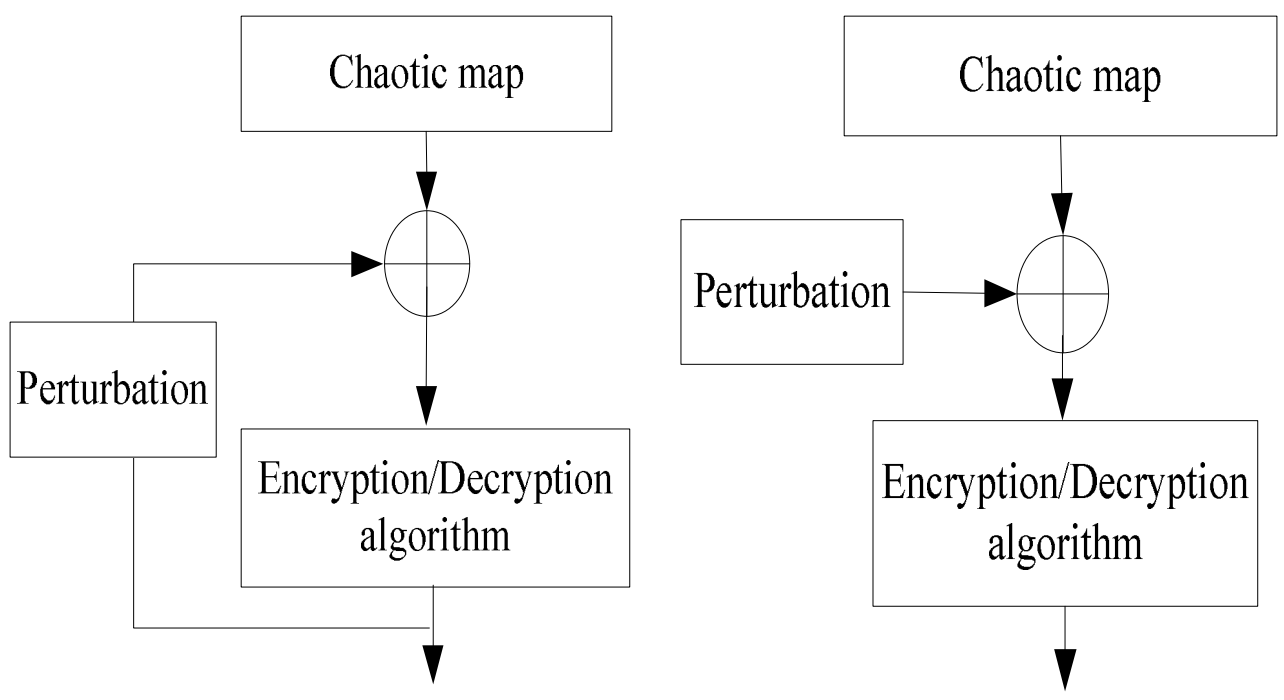

(a) (b)

Fig. 9. Description of the way of perturbation for (a) Socek and Xiang method and (b) the proposed algorithm

The same conclusion can be made for the Lian [28] and Wong algorithms [29]. In the diffusion method proposed by Lian, the pixel value mixing depends on the value of the previously processed pixel. Then, if a transmission error occurs in the encrypted image, we obtain random errors in the decrypted image. In the Wong algorithm, the authors keep the same diffusion process but the confusion method is changed. In their proposition, both the permutation on pixel position and the change of pixel value are carried out in the confusion step. The additional diffusion effect is injected by adding the current pixel value with the previous permuted pixel and then performing a cyclic shift permutation technique. This method also helps to increase the error propagation phenomenon.

We can conclude that the proposed cryptosystem overcomes the drawbacks of existing chaotic algorithms such as the Socek, Yang, Lian and Wong methods.

### 8.2 Encryption time

The encryption times in seconds of different proposed chaotic algorithms and the standard AES in different modes are shown in table 4.

This encryption time is obtained for 4 rounds of the SP box. We can see that the algorithms are faster in OFB operation mode. And, the encryption time obtained with the standard encryption algorithm AES is the highest one. Then, we can conclude that the proposed algorithms are more suitable for image encryption. Between the proposed versions of CBCSTI algorithm, the CBCSTI-E version is the fastest one. It is obvious because this version does not use a 2D chaotic map. The C version using the Arnold map is more rapid than the A version using standard map. But the difference is still negligible (4%) and we will see later that the CBCSTI-A algorithm provides better performance than the E and C versions.

TABLE 4
THE ENCRYPTION TIME IN SECONDS FOR DIFFERENT ALGORITHMS WITH FOUR OPERATION MODES

| | CBC | OFB | CTR | CFB |
|---|---|---|---|---|
| CBCSTI-A | 545 | 525 | 664 | 528 |
| CBCSTI-B | 944 | 921 | 1034 | 932 |
| CBCSTI-C | 527 | 503 | 718 | 525 |
| CBCSTI-D | 956 | 926 | 1050 | 931 |
| CBCSTI-E | 538 | 496 | 644 | 507 |
| AES | 1780 | 1742 | 1806 | 1765 |

### 8.3 NIST Statistical Tests

Among the numerous standard tests for pseudo-randomness, a convincing way to show the randomness of the produced sequences is to confront them to the NIST (National Institute of Standards and Technology) Statistical Tests. The NIST statistical test suite [37] is a statistical package consisting of 188 tests that were developed to test the randomness of arbitrary long binary sequences produced by either hardware or software based cryptographic random or pseudorandom number generators. These tests focus on a variety of different types of non-randomness that could exist in a sequence.

To justify the importance of the 2D chaotic map, we compared the CBCSTI-E, CBCSTI-C and CBCSTI-A methods. Then, we used the above test suite to test the randomness of 100 encrypted images of a length of 2097152 bits. In table 5, we show the results for a number of tests knowing that the sequence passed all the other tests. Note that the 100 encrypted images were generated with randomly selected secret keys. For each test, the default significance level $\alpha$=0.01 was used, at the same time a set of P-values, which corresponding to the set of images, is produced. Each image is deemed a *success* if the corresponding P-value satisfies the condition P-value $\geq \alpha$, and is deemed a *failure* otherwise and noted by a star.

As we can see, the images encrypted with CBCSTI-E methods fail to pass four tests. The CBCSTI-C gives better results but we still have two failed tests. The highest randomness is obtained using the CBCSTI-A method with a minimum of 97% of success.

TABLE 5
NIST STATISTICAL TEST FOR 100 ENCRYPTED IMAGES BY THREE PROPOSED ALGORITHMS CBCSTI-A, D AND E

| | CBCSTI-E | CBCSTI-C | CBCSTI-A |
|---|---|---|---|
| Frequency Monobit Test | 93* | 96* | 97 |
| Block Frequency Test | 99 | 100 | 100 |
| Cumulative Sums Test | 94* | 96* | 98 |
| Random Excursion Test | 95* | 97 | 100 |
| Random Excursion Variant Test | 95* | 98 | 98 |
| Runs Test | 97 | 98 | 98 |
| Longest Runs Test | 97 | 97 | 98 |
| Rank Test | 100 | 100 | 100 |
| Discrete Fourier Transform | 99 | 99 | 100 |
| Serial | 99 | 100 | 100 |
| Non Overlapping Template Matching Test | 98 | 99 | 100 |
| Overlapping Template Matching Test | 99 | 99 | 99 |
| Maurer's « Universal Statistical » Test | 99 | 99 | 100 |
| Linear Complexity Test | 98 | 99 | 99 |

### 8.4 Correlation of two adjacent pixels

Statistical analysis on large amounts of images shows that averagely adjacent 8 to 16 pixels are correlative. To test the correlation between horizontally, vertically and diagonally adjacent pixels from the image, we calculate the correlation coefficient of a sequence of adjacent pixels by using the formulas (24), (25) and (26), (27).

$$E(x) = \frac{1}{MxN}\sum_{i=1}^{M}\sum_{j=1}^{N} P_1(i,j) \tag{24}$$

$$D(P_1) = \frac{1}{MxN}\sum_{i=1}^{M}\sum_{j=1}^{N} [P_1(i,j) - E(P_1(i,j))]^2 \tag{25}$$

$$\mathrm{cov}(P_1, C_1) = \frac{1}{MxN}\sum_{i=1}^{M}\sum_{j=1}^{N} [P_1(i,j) - E(P_1(i,j))][C_1(i,j) - E(C_1(i,j))] \tag{26}$$

$$r_{P_1C_1} = \frac{\mathrm{cov}(P_1, C_1)}{\sqrt{D(P_1)}\sqrt{D(C_1)}} \tag{27}$$

Where $P_1(i,j)$ and $C_1(i,j)$ are gray values of the original pixel and the encrypted one.

Fig.10 shows the correlation distributions of two horizontally adjacent pixels in the first component of the original Mandrill and ciphered images using the CBCSTI-A algorithm. Similar results are obtained for the other components and encryption methods using Mandrill or Lena image.

Firstly, we randomly select 250000 pairs of two horizontally adjacent pixels from the image. Then, we plot the pixel value on location *(x+1,y)* over the pixel value on location *(x, y)*. This test is done for the first color component and similar results are obtained for the Green and Blue components.

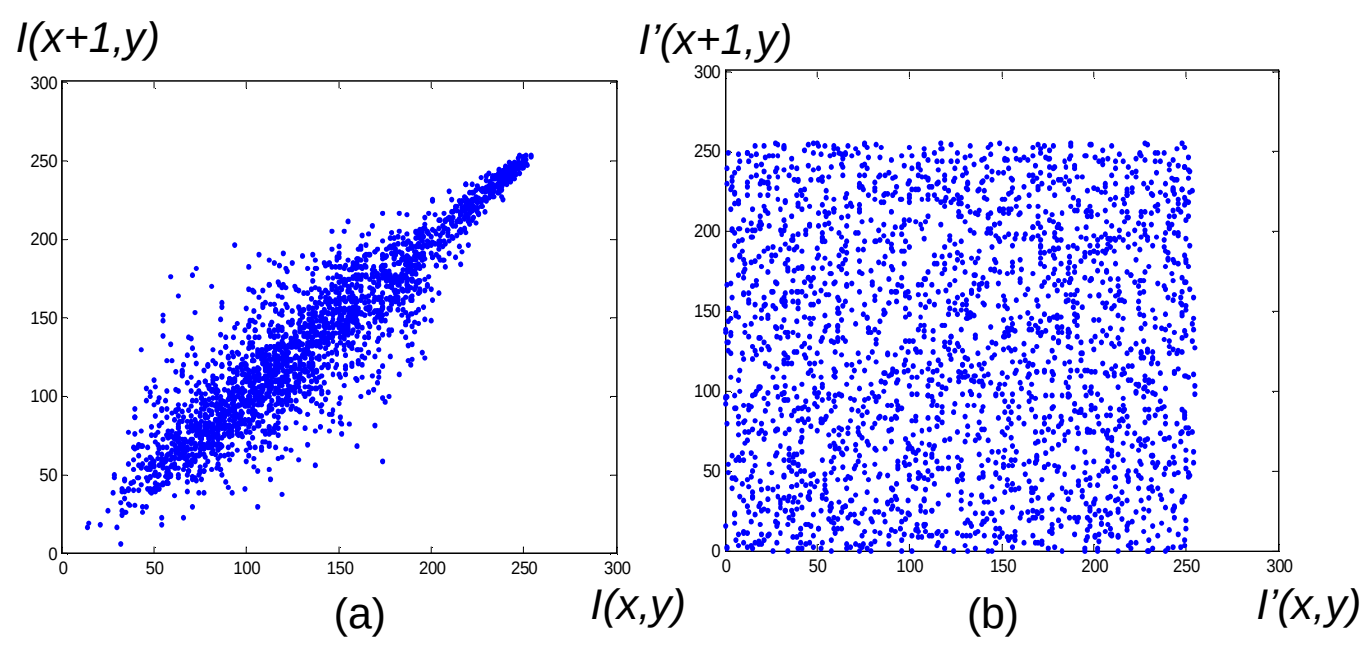

Fig.10. The correlation distributions of two horizontally adjacent pixels in the (a) original image and (b) ciphered image.

In the last section 8.3, we verified the impact of the 2D chaotic map using NIST Tests. In this section, we compare the contribution of Standard and Arnold maps using the correlation of adjacent pixels as indicator.

In table 6, the mean values of the correlation coefficients are shown for the original Mandrill and Lena images. Table 7 gives the coefficients for the permuted images of Mandrill by Arnold and standard maps for a number of iterations It = 3 and 5. And finally, in Table 8, we present the coefficients of the encrypted images by CBCSTI-A, CBCSTI-E and the standard AES in OFB operation mode.

TABLE 6

MEAN VALUES OF THE CORRELATION COEFFICIENTS OF ADJACENT PIXELS IN THE ORIGINAL MANDRILL AND LENA IMAGES

| | Mandrill | Lena |
|---|---|---|
| Horizontal correlation | 0.9203 | 0.9829 |
| Vertical correlation | 0.8631 | 0.9907 |
| Diagonal correlation | 0.8494 | 0.9722 |

TABLE 7
MEAN VALUES OF THE CORRELATION COEFFICIENTS OF ADJACENT PIXELS IN THE PERMUTED IMAGES OF MANDRILL BY ARNOLD AND STANDARD MAPS.

| Correlation | Arnold map | | Standard map | |
|---|---|---|---|---|
| | It=3 | It=5 | It=3 | It=5 |
| horizontal | 0.7056 | 0.3166 | 0.1293 | 0 .0176 |
| vertical | 0.6593 | 0 .0767 | 0.0113 | -0.0082 |
| diagonal | 0.7519 | 0.4684 | 0.0096 | -0.0254 |

As we can see in Table 7, the correlation coefficients are inversely proportional to the number of iterations. When the number of iterations increases, the correlation values are lower.

We will see later that in CBCSTI algorithms, a number of iterations It=3 is good enough to obtain low correlation coefficients. Moreover, the correlation coefficients obtained in the case of the Standard map are significantly lower than the obtained values for the Arnold map. Similar results are obtained for Lena image. For example, for It=3, we find 0.1888 as horizontal correlation, 0.0043 as vertical correlation and 0.0559 as the diagonal correlation.

Despite the good correlation results, it is obvious that the use of a 2D chaotic map alone is not sufficient to do the encryption. 2D chaotic map permutes only the position of pixels and does not change their values. Then, the original and permuted images have the same statistical characteristics as the histogram and the entropy value. Then, the cryptanalysis in this case is not very complicated.

Table 8 presents the mean values of the correlation coefficients of adjacent pixels in the encrypted images by CBCSTI-C, CBCSTI-A and AES encryption algorithms in OFB operation mode. Similar results are obtained for CBC, CFB and CTR modes.

TABLE 8
MEAN VALUES OF THE CORRELATION COEFFICIENTS OF ADJACENT PIXELS FOR CBCSTI-A, E AND AES ALGORITHMS.

| Correlation | Algorithm | | | |
|---|---|---|---|---|
| | Mandrill | | Lena | |
| | AES | CBCSTI-C | CBCSTI-A | CBCSTI-A |
| horizontal | -0.0188 | -0.0064 | -0.0092 | 0.0127 |
| vertical | -0.0024 | -0.0126 | 0.0096 | 0.0093 |
| diagonal | 0.0019 | 0.0374 | -0.0083 | -0.0059 |

It can be seen that the use of the standard chaotic map in CBCSTI-A version significantly reduces the vertical and diagonal correlation coefficients with a minimum of 66% compared to the CBCSTI-C method. The horizontal correlation is slightly higher but still very low. The AES algorithm gives lower vertical and diagonal correlation coefficients but the horizontal coefficient is still higher than that obtained with the CBCSTI-A method. We can observe that the performance of the proposed scheme is roughly equivalent to AES and that the Standard method remains better than the Arnold map.

## 8.5 Information entropy analysis

Entropy is a statistical measure of randomness that can be used to characterize the texture of an image. It is well known that the entropy *H(m)* of a message source m can be calculated as [30]:

$$H(m) = \sum_{i=0}^{2^N-1} p(m_i)\log_2 \frac{1}{p(m_i)} \qquad (28)$$

Where $p(m_i)$ represents the probability of message $m_i$.

When an image is encrypted, its entropy should ideally be 8. If it is less than this value, a certain degree of predictability exists which threatens its security.

In table 9, we list the entropy of the images encrypted of Mandrill by three algorithms, CBCSTI-A, CBCSTI-C and AES in OFB operation mode. Similar results are obtained for CBC, CFB and CTR modes and using Lena image.

TABLE 9
ENTROPY VALUE FOR THE IMAGES ENCRYPTED WITH THREE DIFFERENT ALGORITHMS

| Algorithm | Original image | CBCSTI-A | CBCSTI-C | AES |
|---|---|---|---|---|
| entropy | 7.7624 | 7.9999 | 7.9997 | 7.9993 |

The value obtained for CBCSTI-A is the nearest to the theoretical value 8. This means that information leakage in the encryption process is negligible and the encryption system is secure against entropy attack. Lower entropy is obtained using the AES method but it is also very close to 8.

## 8.6 Comparison between original and encrypted image

**Histogram analysis**

An image-histogram illustrates how pixels in an image are distributed by graphing the number of pixels at each color intensity level. We have calculated and analyzed the histograms of the encrypted image as well as the original colored image. Fig. 11 shows the histogram of the Red component of the original Mandrill image as well as images ciphered using the CBCSTI-A algorithm in OFB operation mode. Similar results are obtained for the other color components? the proposed encryption methods in different operation modes and for Lena image.

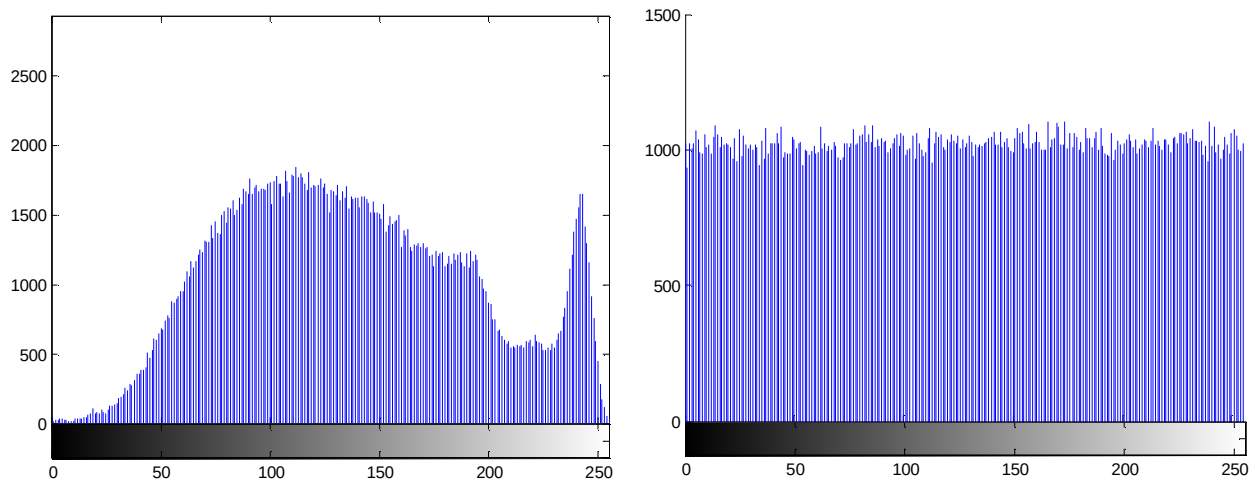

(a) (b)

Fig.11. Histograms of red component of (a) 'MANDRILL.BMP' image and (b) the encrypted image.

As we can see, the histogram of the ciphered image is fairly uniform and is significantly different from that of the original image of Mandrill.

**Difference between the original and the ciphered images**

Common measures like correlation, NPCR (Number of pixels change rate) and UACI (Unified Average Changing Intensity) are used to test the difference between the original image *P1* and the encrypted one *C1*.

We calculate the correlation coefficient r of the original and encrypted images by using the formulas (24), (25), (26) and(27).

NPCR stands for the number of pixel change rate. Then, if D is a matrix with the same size as images $P_1$ and $C_1$, $D(i,j)$ is determined as follows (29):

$$D(i,j)=\begin{cases}1 & if\ \ P_1(i,j)\neq C_1(i,j)\\ 0 & else\end{cases} \tag{29}$$

NPCR is defined by the following formula (30):

$$NPCR=\frac{\sum_{i=0}^{M-1}\sum_{j=0}^{N-1}D(i,j)}{M\times N}\times 100 \tag{30}$$

Where, $M$ and $N$ are the width and height of $P_1$ and $C_1$.

The UACI measures the average intensity of differences between the plain image and the ciphered image. UACI is defined by the following formula (31):

$$UACI=\frac{1}{MxN}\sum_{i=0}^{M-1}\sum_{j=0}^{N-1}\frac{|P_1(i,j)-C_1(i,j)|}{255}x100 \tag{31}$$

In table 10, we summarize the Mean value of the correlation, NPCR and UACI obtained between the original image and the encrypted one with the CBCSTI-A algorithm in OFB operation mode when the three color components are considered. Similar results are obtained in the three other operation modes.

TABLE 10
MEAN VALUE OF THE CORRELATION, NPCR AND UACI BETWEEN THE ORIGINAL IMAGES AND THE ENCRYPTED ONES

| Image | Correlation | NPCR(%) | UACI (%) |
|---|---|---|---|
| Mandrill | 0,0019 | 99.61 | 29.94 |
| Lena | 0,0014 | 99.62 | 30.42 |

We can see that we have obtained a low correlation between the original and the ciphered image. The NPCR and UACI are high enough to say that the two images are very different.

### 8.7 key sensitivity

An encryption scheme has to be key-sensitive, meaning that a tiny change in the key will cause a significant change in the output. In order to demonstrate the key sensitivity of the CBCSTI-A algorithm in OFB operation mode, the following experiments have been carried out with a slightly different key.

**Key sensitivity at the emission**

Fig.8 shows the encrypted image with the following key:

alpha= 0.35899926, beta=0.25899926, $x_0$=0.7239 and $y_0$= 0.5672. alpha and beta are the control parameters of the chaotic maps and $x_0$, $y_0$ are their initial values. We encrypt the same image using the slightly changed key as follows: alpha= 0.3589992600001. Then, Fig. 12 shows the difference between the two ciphered images where the three color components are considered.

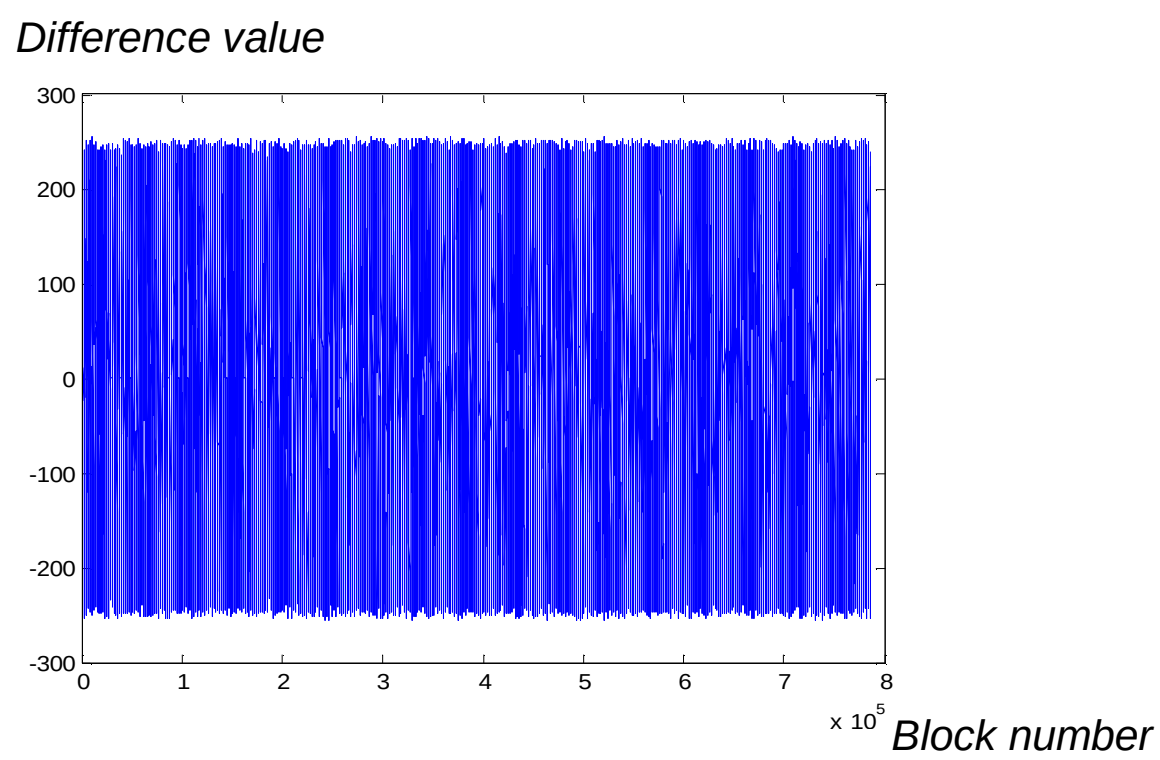

Fig.12. The difference between two ciphered images with a very small changed key.

As we can see even where the control parameter of the first orbit is changed slightly ($10^{-8}$), the encrypted image is absolutely different from the first one. This difference is of 53% different pixels for the first color component, 99% for the second one and 100% for the last one.

We can observe that the difference increases between the two images when passing from one component to another. The reason lies in the behaviour of the chaotic orbit. In fact, we begin with close control parameters. Then, with iterations, we obtain two divergent chaotic orbits and then two different encrypted images.

The comparison of changed intensities leads to the same conclusion with a mean value of 30%.

This test is also performed for the algorithm in the other operation modes but OFB operation mode gives the highest performance. For example, in CBC mode, we obtain a mean value of 71% of different pixels with 24% of changing intensity. Similar results are obtained for Lena image.

### 8.8 Plaintext sensitivity

This test compares the sensitivity of our proposed encryption method to the plaintext. To carry out this test, we encrypt two plain images of Mandrill having only a 1 bit difference at the upper left corner. Then, we measure the difference between the two ciphered images using the CBCSTI-A method in different operation modes.

In CBC mode, the two ciphered images have 100% different pixels with a rate of change of 27% and similar results are obtained for CFB mode.

In OFB and CTR modes, we have one changed block between the two encrypted images.

Although OFB mode presents some advantages in key sensitivity tests and avoids the propagation of errors in the decrypted image, its robustness to the chosen plaintext attack and differential attack is not guaranteed. But it is relative to the chosen operation mode and not the proposed encryption algorithm. CBC mode doubles the errors in decrypted images but the algorithm in this mode seems more resistant to this type of cryptanalysis.

## 9 CONCLUSION

In this paper, a new chaos-based cryptosystem is proposed. Our cryptosystem shares some similarities with those proposed by Socek, Lian and Wong, but produces cryptograms suitable to be transmitted on insecure and noisy channels. Indeed, in the new encryption/decryption algorithms, the key space is larger and the multi-rounds S-P network operations on each pixel are more complex than some existing algorithms. The proposed SP box including the pixel permutation method enhanced the statistical properties of the encrypted images. Furthermore, the introduction of the perturbation technique with a LFSR and the architecture of the algorithm allow us to avoid the propagation of errors in the decrypted images when a transmission error occurs. A study of the algorithm in different operation modes is shown and the obtained results: uniformity, key sensitivity, correlation, entropy, NIST statistical tests, proves the high cryptographic quality of the proposed cryptosystem and its compliance with NIST standards.